\documentclass[10pt,times,twocolumn]{article}

\usepackage{sbcm2019}
\usepackage{graphicx,url}

\usepackage{anonymize} 

 \usepackage[utf8]{inputenc}

\title{\anonymize{Batebit} Controller: Popularizing Digital Musical Instruments' Development Process}

\author{
\anonymize{Filipe Calegario}\inst{1} \inst{2}
\and
\anonymize{João Tragtenberg}\inst{1} \inst{2}
\and
\anonymize{Giordano Cabral}\inst{1}
\and
\anonymize{Geber Ramalho}\inst{1}
}

\address{\anonymize{MusTIC / Centro de Informática / Universidade Federal de Pernambuco}\\
         \anonymize{Av. Jornalista Aníbal Fernandes, s/n - Cidade Universitária (Campus Recife) - 50.740-560 - Recife - PE}
         \nextinstitute
         \anonymize{CIIMUS / Instituto SENAI de Inovação para TICs / SENAI-FIEPE} \\
         \anonymize{Rua Frei Cassimiro, 88 - Santo Amaro - 50.100-260 - Recife - PE}
         \email{\anonymize{fcac@cin.ufpe.br, tragtenberg@gmail.com, grec@cin.ufpe.br, glr@cin.ufpe.br}}
}

\begin{document}

\maketitle

\begin{abstract}
In this paper, we present an ongoing research project related to popularizing the mindset of building new digital musical instruments. We developed a physical kit and software intended to provide beginner users with the first grasp on the development process of a digital musical instrument. We expect that, by using the kit and the software, the users could experiment in a short period the various steps in developing a DMI such as physical structure, electronics, programming, mapping, and sound design. Our approach to popularizing the DMI development process is twofold: reducing the cognitive load for beginners by encapsulating technical details and lowering the costs of the kit by using simple components and open-source software. In the end, we expect that by increasing the interest of beginners in the building process of digital musical instruments, we could make the community of new interfaces for musical expression stronger.
\end{abstract}

\section{Introduction}

This paper focuses on the technical details of the development process of a digital musical instrument. We believe that overcoming technical barriers by presenting them in a natural approach could open up possibilities for beginners. In sum, we hypothesize that when the technical barrier is presented as a natural step, the beginners can open their heads for new ideas.

The development of new digital musical instruments (DMIs) is an interdisciplinary process, in which each step relates to a specific mindset and requires a particular set of skills \cite{DalFarra2018}. Mechanical structure, electronics, programming, mapping, and sound design can be loosely considered the development areas of a DMI.

Due to these different mindsets and skills, developing a DMI can become a laborious process. In this sense, the designer must manage different partnerships or become a polymath to achieve the musical artifact at the end.

Following the example of Arduino \cite{Banzi2009}, which popularized the physical computing by simplifying the access of artists and designers into electronic prototyping, we believe that more straightforward and faster ways of developing new DMIs can contribute to more beginners experimenting ideas and, therefore, engagement in the DMI community.

By understanding the process, the users can adapt the technology to their needs, intentions, and contexts of use. The more people are experimenting and testing, the higher the chances of achieving great instruments ideas.

In this paper, we present a physical kit and software to serve as an entry point to DMI development. Our approach is to reduce cognitive load by encapsulating technical details. Besides, we consider that it is essential to make the kit accessible to broadening the audience. Therefore, we propose to use accessible and straightforward components, materials and manufacturing techniques, and open-source software, in order to reduce the costs and reach more people.

\section{Related Projects}

There is a considerable amount of hardware and software tools already available that could be used during DMI development process. Some examples are: microcontrollers environments (Arduino, Raspberry Pi, Beaglebone, Teensy), sensor kits (Infusion Systems, littlebits, makey makey), MIDI controllers (keyboard, wind controllers, percussion controllers), general-purpose programming languages (C, C++, Java), audio-oriented programming languages (CSound, SuperCollider, Chuck, Pure Data, Max/MSP), creative programming environments (Processing, openframeworks, Cinder, Scratch), applications for mappings (libmapper, iCon, OSCulator, juxion, Wekinator), and digital audio workstations (Logic Pro, Ableton Live, Pro Tools, GarageBand, Reaktor, Tassman).

Beyond those tools, there is a growing number of DMI's development toolkits aimed to reduce the technical barriers for musicians and designers \cite{McPherson2017, Armitage2018, Calegario2017}. However, we believe that these toolkits still presents an in-depth approach concerning the aspects of DMI development. For instance, Bela \footnote{http://bela.io}, a platform for musical interactions based on the BeagleBone board, focus on providing real-time processes and more natural ways of programming musical interactions. There is a significant number of examples that can help the users begin to develop their instruments, but they have to build their physical interface, mechanical structure, and electronics. There is no doubt that, with Bela, the users could have a profound, expressive result. However, for the popularization of DMI development, we find that it is more interesting to have an in-breadth approach.

\section{Our approach}

Concerning the beginners, we believe that an in-breadth approach contributes to a better understanding of the DMI development process. In other words, didactically, we consider that allowing the users to experiment a small, superficial amount of each stage of DMI technical development in a short period can make them better comprehend the process and identify what aspect excites them.

We intend to embed a little portion of the DMI development process into an object so the users can have an enactive experience when using it. We believe that the feeling of making an instrument with their own hands from zero to a functional prototype in a short period with little pitfalls is a productive way of teaching the complex and arduous process of developing a DMI.

\begin{figure}[htb]
    \centering
        \includegraphics[width=0.78\columnwidth]{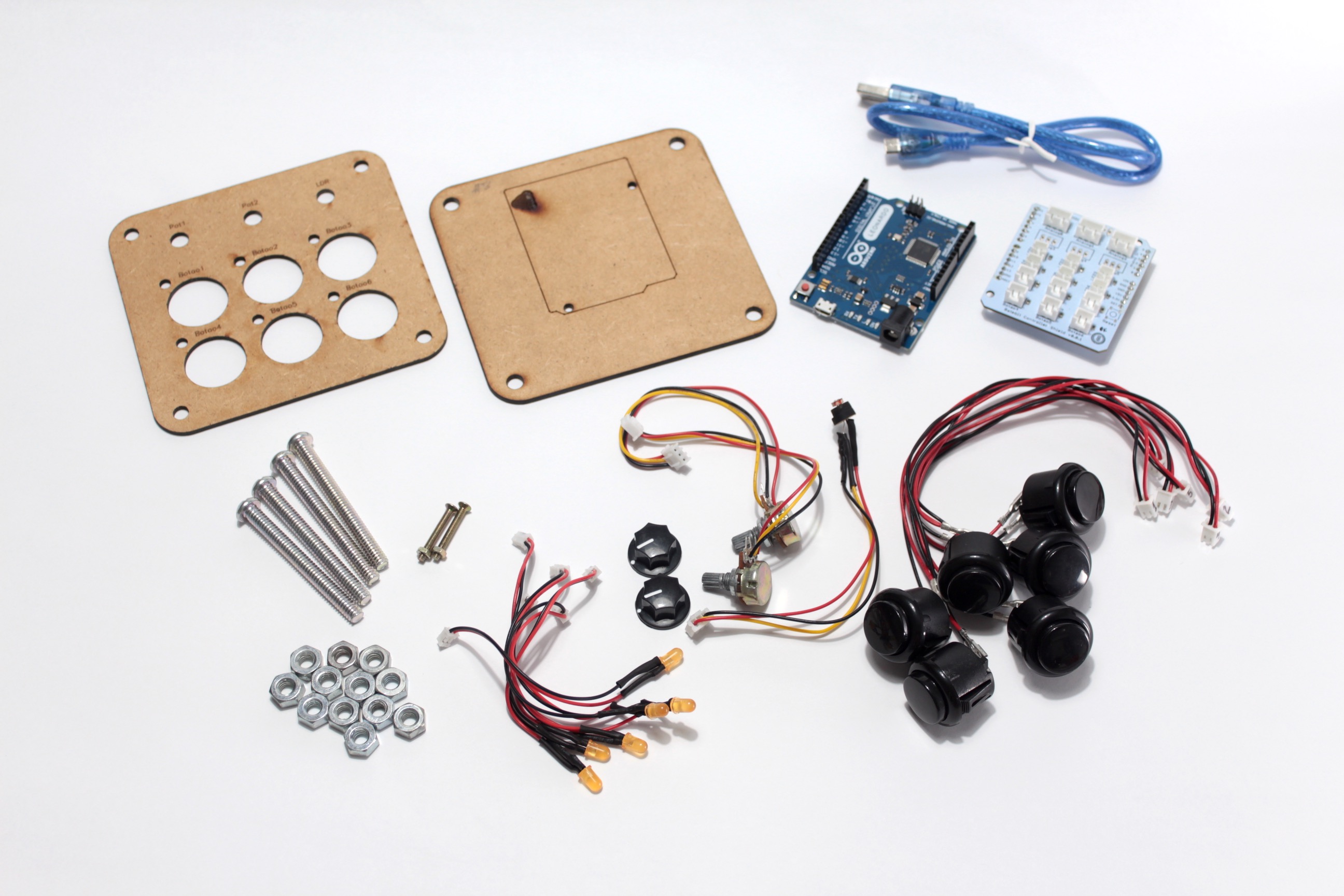}
    \caption{Disassembled Kit}
    \label{fig:kitDesmontado}
\end{figure}

The idea is to guide the users throughout the mechanical structure with MDF plates, screws and bolts; the electronics with a Arduino Leonardo, a shield, two knobs, one LDR (light sensor), six buttons, and six LEDs; the programming with a example code in Arduino IDE; the mapping and sound design with a patch in Pure Data. Figure \ref{fig:kitDesmontado} shows the disassembled kit.

\begin{figure}[htb]
    \centering
        \includegraphics[width=0.78\columnwidth]{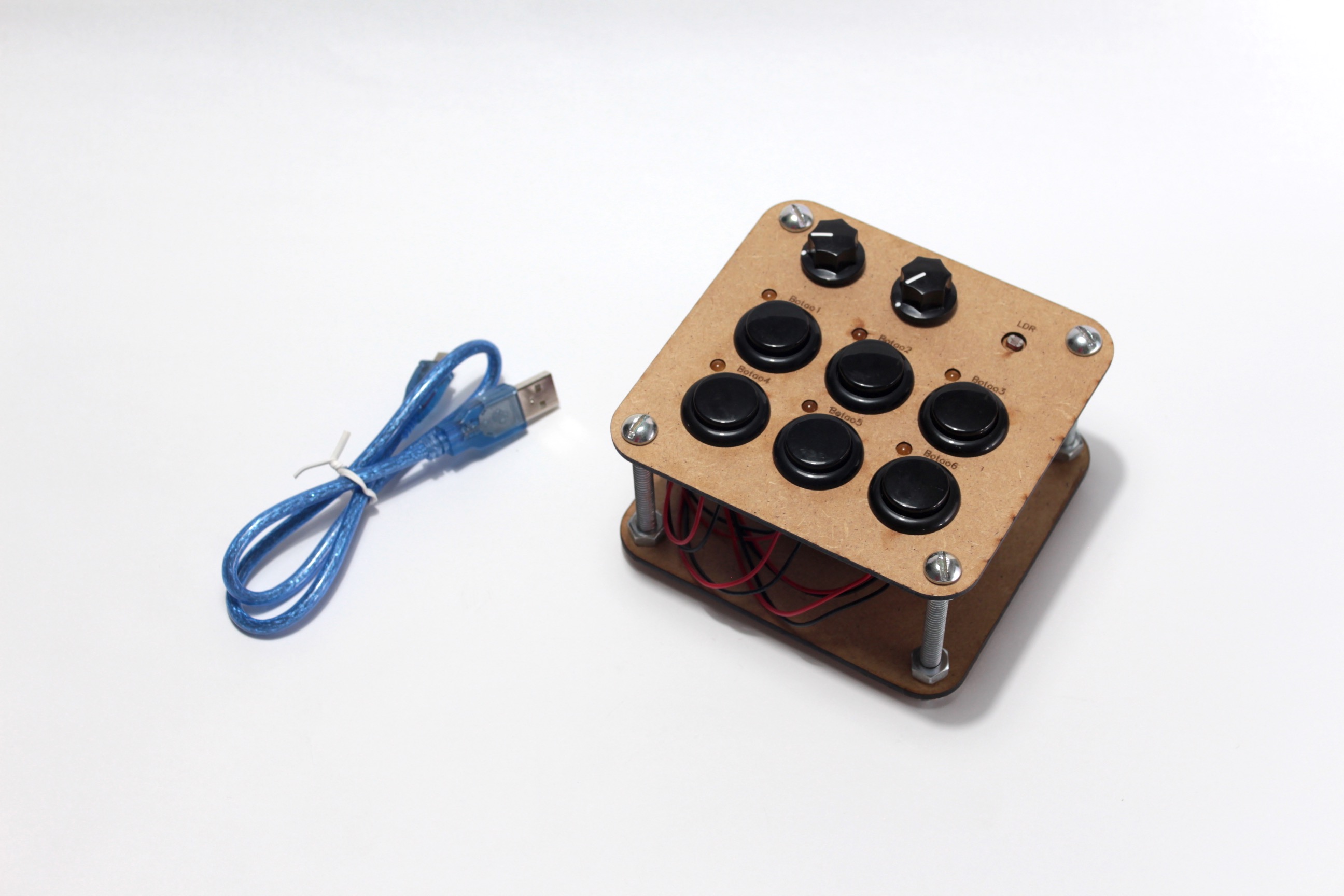}
    \caption{Assembled Kit}
    \label{fig:kitMontado}
\end{figure}

When assembled (Figure \ref{fig:kitMontado}) and connected to the computer via USB cable, the kit becomes a MIDI controller that operates the parameters of a Pure Data patch. The buttons trigger samples and light of the LEDs. The two knobs change the time interval and the feedback amount of a delay effect. The LDR is related to the pitch parameter of an oscillator, which is gated by a dedicated button. 

\section{Workshops}

The kit and software were informally tested in three workshops. Each one had twenty participants that worked in pairs. The kits were presented disassembled, and the participants had to follow a set of instructions in a PDF illustrated with photos. The pairs were encouraged to work without waiting for guidance from the workshop facilitator, that should be called only in the occurrence of an obvious error.

In two hours, the participants had to assemble, program, and play the \anonymize{Batebit} Controller. After that, the participants answered a questionnaire with multiple choices and open text questions. One participant mentioned that he was expecting to see, but it was not covered the possibility to have a final product, instead of a prototype: \textit{``How the project ceases to be a prototype and becomes a final product that I can take it to the stage?''}(P09). Moreover, two participants mentioned that the most critical aspect of the workshop for them was to have an overview of the entire process: \textit{``The [most relevant part was] overview of what is happening in the entire process''}(P08) and \textit{``The step-by-step of the process''} (P06). 

\section{Conclusions}
The \anonymize{Batebit} Controller is a simple in-breadth approach to embed the DMI technical development process into a kit and software that can be didactically experienced in a short period by beginners. The objectives are to be the entry point for popularizing the mindset of instrument making, reach a broader audience, and make DMI community stronger. The feedback of the workshop participants showed that the kit and software have the potential to inspired non-technical users to begin experimenting with DMI development. In future steps of the research, we plan to better understand the engagement of the participants on each phase of the technical development process.

\bibliographystyle{unsrt}
\bibliography{nime-references}

\end{document}